\documentclass[twoside,leqno,twocolumn]{article}
\usepackage{ltexpprt}
\usepackage{amsmath}
\usepackage{amssymb}
\usepackage{graphicx}
\begin{document}

\title{\Large  Data-Driven Koopman Analysis of Tropical Climate Space-Time Variability}
\author{Joanna Slawinska\thanks{Center for Environmental Prediction, Rutgers University, joanna.slawinska@nyu.edu},
Eniko Sz\'ekely\thanks{Courant Institute of Mathematical Science, New York University},
Dimitrios Giannakis\thanks{Courant Institute of Mathematical Science, New York University}}
\date{}

\maketitle

\pagenumbering{arabic}
\setcounter{page}{1}

\begin{abstract} \small\baselineskip=9pt
We study nonlinear dynamics of the Earth's tropical climate system. For that, we apply a recently developed technique for feature extraction and mode decomposition of spatiotemporal data generated by ergodic dynamical systems. The method relies on constructing low-dimensional representations (temporal patterns) of signals using eigenfunctions of Koopman operators governing the evolution of observables in ergodic dynamical systems. We apply this technique to a variety of tropical climate datasets and extract a multiscale hierarchy of spatiotemporal patterns on diurnal to interannual timescales. In particular, we detect without prefiltering the input data modes operating on intraseasonal and shorter timescales that correspond to propagation of organized convection. We discuss the salient properties of these propagating features and in particular we focus on how the activity of certain types of these traveling patterns is related to lower-frequency dynamics. As an extension of this work, we discuss their potential predictability based on a range of nonparametric techniques and potential advances related to understanding the deterministic and stochastic aspects of the variability of these modes.

\end{abstract}

\section{Introduction}

The Earth's tropical climate system consists of many components that evolve in time over convective to planetary scales and are coupled with each other. For example, the tropical atmosphere of the Indo-Pacific basin incorporates the Walker cell \cite{LAU2015}, which consists of a planetary-scale horizontal flow (zonal extent of order $10^4$ km) and vertical motions embedded within the troposphere (depth of order 10 km). Moreover, the specific spatial structure of the Walker cell is determined by its boundary (oceanic) conditions, which are spatially inhomogeneous. In particular, the Western Pacific `Warm Pool' is climatologically characterized by higher sea surface temperature (SST), roughly 10 K more than over the eastern tropical Pacific basin that constitutes the eastern edge of the Walker cell ($\sim 100^\circ$E longitude).  Importantly, tropical Pacific SST fluctuates around its climatological values over many timescales. The most pronounced anomalies reach an amplitude of 3--4 K and are observed quasi-periodically every 3-8 years.  It has been recognized that atmosphere-ocean couplings \cite{BJERKNES69, WYRTKI1975, CANEZEBIAK1985}, are at the heart of this interannual variability, known commonly as the El Ni\~no-Southern Oscillation (ENSO).

The atmospheric component of the tropical climate consists predominantly of convective processes \cite{EMANUEL1995}. In particular, individual convective events are triggered by turbulent plumes of moist air rising intermittently from the atmospheric boundary layer, which is heated from below by warm oceanic waters. Such localized perturbations can evolve into convective cells varying greatly in size and shape, with some reaching the spatial scale of the tropospheric depth (tens of km) and lasting for a few hours. These individual clouds arrange themselves spatially into progressively larger structures from mesoscale, to synoptic, and planetary scales \cite{MAPES1993, MAPES2006}.  Frequently, numerous organized convective systems propagate zonally \cite{SLAWINSKA2009}, some of them being associated with convectively coupled equatorial waves (CCEWs)\cite{KILADIS2009}, or with intraseasonal oscillations such as the Madden-Julian oscillation (MJO) \cite{MADDEN1972, LAU2011}. The temporal evolution of all scales depends crucially on convective scale processes that, through various positive and negative feedbacks and with different temporal lags, enhance and suppress fluctuations associated with a given scale  \cite{SLAWINSKA2014}.

Understanding the Earth's tropical climate requires better understanding of underlying nonlinearities. Classical linear techniques, such as empirical orthogonal function (EOF) analysis, typically require filtering or detrending of the data to isolate the timescale of interest and thus have a risk of extracting modes of limited physical significance. In this work, we demonstrate the potential of recently introduced methods in decomposing high-dimensional time series into dominant patterns associated with its dynamical evolution. These methods, introduced briefly in section 2, take advantage of the framework of ergodic theory and are implemented using  machine learning tools (e.g., kernel methods). Unlike classical approaches,  they allow for simultaneous extraction of  many timescales with no ad hoc preprocessing of the input data. Section 3 demonstrates the utility of these methods applied to the Earth's tropical climate, and section~4 discussed examples of important physical questions that can be addressed in future work.

\section{Data-driven spectral methods for dynamical systems}

\subsection{Introduction to Koopman operators}

Our approach for extracting coherent spatiotemporal patterns combines aspects of spectral theory of dynamical systems \cite{BerryEtAl15,GiannakisEtAl15,Giannakis16,DasGiannakis17} with kernel algorithms for machine learning \cite{GiannakisMajda11c,GiannakisMajda12a,Giannakis15}. In particular, consider a dataset consisting of time-ordered observations $ x_0, x_1, \ldots, x_{N-1} $ of a climatic variable taken at times $ t_0, t_1, \ldots, t_{N-1} $ with $ t_n = n\tau $ for a fixed sampling interval $ \tau $. In the applications studied here, $ x_n \in \mathbb{ R }^d $ corresponds to cloud top temperatures measurements over tropical band sampled at $ d $ gridpoints. Here, we view the $ x_n $ as the values of an observable, i.e., a function of the state, of an abstract ergodic dynamical system (in this case, the Earth's  tropical climate system); that is, we consider that there exists a mapping $ F : A \mapsto \mathbb{ R }^d $ from the state space $ A $ of the dynamical system such that $ x_n = F( a_n ) $, where $ a_n \in A $ is the state at time $ t_n $. Moreover, the dynamics on $ A $ are described by a (generally nonlinear) mapping $ \Phi_t : A \mapsto A $, $ t \in \mathbb{ R } $, such that $ a_n = \Phi_{n\tau}( a_0 ) $ for an initial state $ a_0 $. That mapping is assumed to be ergodic for an invariant probability measure $ \alpha $; in particular, for any observable $ f \in L^1( A, \alpha ) $ and $ \alpha $-a.e.\ starting state $ a_0 $ we have 
\begin{equation}
  \label{eqErgodicity}
  \lim_{N\to\infty} \frac{ 1 }{ N } \sum_{n=0}^{N-1} f( a_n ) = \int_A f \, d\alpha.
\end{equation}
Ergodicity is a key property that allows the data-driven implementation of our techniques. Note that besides this property we do not assume that we have a priori knowledge of the state space $ A $ or the dynamical flow $ \Phi_t $.

Unlike traditional modeling approaches based on the state-space representation of dynamical systems, the spectral theory of dynamical system is based on an equivalent representation involving the action of the dynamics on spaces of observables. For our purposes, it suffices to consider the Hilbert space $ H = L^2(A,\alpha) $ of complex-valued square-integrable functions on $ A $ with respect to the invariant measure $ \alpha $, equipped with the inner product $ \langle f_1, f_2 \rangle = \int_A f^*_1 f_2 \, d\alpha $. Note that this inner product can be approximated from time series of the values of $ f_1 $ and $ f_2 $ using~\eqref{eqErgodicity} with $ f = f^*_1 f_2 $. On $ H $, the dynamical system acts via Koopman operators, that is, unitary operators $ U_t : H \mapsto H $, $ t \in \mathbb{ R } $, defined by $ U_t f = f \circ \Phi_t $. These operators form a group with $ U_t U_s = U_{t+s}$. A key property of this representation is that the space $ H $ and the operators $ U_t $ are \emph{linear} (without approximation) even if $ A $ and $\Phi_t $ are nonlinear. In effect, we are trading off a nonlinear dynamical evolution law ($\Phi_t$)  on a finite-dimensional state space ($A$) by a group of linear operators ($U_t$) on an infinite-dimensional linear space of observables ($H$). 

While the fact that $ H $ is infinite-dimensional might appear at first as a significant drawback, this space and the corresponding Koopman operators $ U_t $ are amenable to study and approximation from data using the full machinery of functional analysis, ergodic theory, and harmonic analysis, and are useful for carrying out tasks such as pattern extraction and prediction. Historically, this line of research was initiated in \cite{DellnitzJunge99,MezicBanaszuk04,Mezic05,Schmid10}, and has evolved into a rich area with many domain applications in science and engineering (e.g., \cite{BudisicEtAl12} and references therein). The papers \cite{BerryEtAl15,GiannakisEtAl15,Giannakis16,DasGiannakis17} have continued on this program by leveraging kernel algorithms for machine learning to enable efficient and robust approximation of Koopman and related operators from high-dimensional time series with rigorous convergence guarantees. 

In this work, we take advantage of the observation made in \cite{Giannakis16,DasGiannakis17} that a class of kernel operators (defined in a suitable space of delay-embedded data as described below), commutes, and therefore has common eigenfunctions, with the Koopman operator $ U_t $. This is a particularly desirable property for the purpose of pattern extraction since it is a standard property of Koopman eigenfunctions of measure preserving systems that they evolve periodically in time, even if the dynamical system itself is non-periodic. In particular, if $ z \in H $ is an eigenfunction of $ U_t $, we have 
\begin{equation}
  \label{eqUTEig}
  U_t z = e^{i \omega t} z, 
\end{equation}
where $ \omega \in \mathbb{ R } $ is an intrinsic frequency of the dynamical system. Thus, one can view Koopman eigenfunction decomposition as a generalized Fourier analysis for a frequency spectrum that is intrinsic to the dynamical system generating the data. Typically, for sufficiently complex (mixing) systems $ U_t $ will also have continuous spectrum, and for this part of the spectrum a notion of approximate Koopman eigenfunctions can also be defined by adding a small amount of diffusion \cite{Giannakis16}, yielding quasiperiodic observables with high temporal coherence. Another advantage of Koopman eigenfunctions is that they do not depend on the observation function $ F $; this enables fusion of data acquired from different sensors in terms of a universal set of coordinates obtained from Koopman eigenfunctions. 

\subsection{Data-driven approximation}

We now describe our approach for computing Koopman eigenfunctions from the time-ordered observations $ x_1, \ldots, x_{N-1} $. First, we embed the input data into a higher-dimensional space using Takens' method of delays \cite{SauerEtAl91}, viz. 
\begin{displaymath}
  x_n \mapsto y_n = ( x_n, x_{n-1}, \ldots, x_{n-q+1} ) \in \mathbb{R }^{qd},
\end{displaymath}
where $ q $ is the number of delays. Note that the time series $ y_0, y_1, \ldots, y_{N-1} $ obtained in this way can also be described in terms of an observable $ F_q : A \mapsto \mathbb{ R}^{qd} $ of the dynamical system such that $ y_n = F_q( a_n )$. Second, we introduce a kernel function $ K: \mathbb{ R }^{qd} \times \mathbb{R}^{qd} \mapsto \mathbb{R}_+ $ that operates on data in delay-embedding space. A wide class of kernels could be used for that purpose, but in what follows we work with the class of variable-bandwidth kernels introduced in \cite{BerryHarlim16}, 
\begin{equation}
  \label{eqK}
  K( y_m, y_n ) = \exp\left( - \frac{ \lVert y_m-  y_n\rVert^2 }{ \epsilon r^{-1/{d_A}}( y_m ) r^{-1/{d_A}}( y_n ) }\right),
\end{equation}
where $ r( y_m ) = \sum_{n=0}^{N-1} e^{-\lVert y_m - y_n \rVert^2 / \epsilon } $ and $ d_A $ are estimates of the sampling density of the data in delay-embedding space and the dimension of $ A $, respectively; see \cite{BerryEtAl15} for details. Note that this bandwidth function is important for ensuring orthogonality of the kernel eigenfunctions with respect to the invariant measure $\alpha$ of the dynamics. Note also that the kernel in~\eqref{eqK} implicitly pulls back to a kernel $ K_A : A \times A \mapsto \mathbb{ R }_+$ on the state space such that $ K_A( a_m, a_n ) = K( y_m, y_n ) $. 

Next, we define a kernel integral operator $ G: \mathbb{ R }^N \mapsto \mathbb{ R}^N $ acting on observables $ f $ of the dynamical system sampled at the  $N $ states;  in particular, we set
\begin{displaymath}
  G f = \frac{ 1 }{ N } \sum_{n=0}^{N-1} K( \cdot, a_n ) f( a_n ). 
\end{displaymath}
By ergodicity, $ G $ approximates the action of an integral operator $ \mathcal{ G } : H \mapsto H $ on the Hilbert space associated with the invariant measure of the dynamics; that is, according to~\eqref{eqErgodicity}, 
\begin{displaymath}
  G f \stackrel{\text{a.s.}}{\longrightarrow}  \mathcal{ G} f = \int_A K_A( \cdot, a) f(a)\, d\alpha(a).
\end{displaymath}
Finally, we normalize $ G $ to a Markov operator $ P : \mathbb{ R }^N \to \mathbb{ R}^N $ by applying the kernel normalization procedure introduced in the diffusion maps algorithm \cite{CoifmanLafon06}, and compute the eigenvalues and eigenfunctions of this operator, $ P \vec \phi_k = \lambda_k \vec \phi_k $, $ k \in \{ 0, 1, \ldots, N -1 \} $. By convention, we order the eigenvalues in non-increasing order  (the top eigenvalue $ \lambda_0 $ is always equal to 1 since $ P $ is Markov).  

The eigenfunctions $ \vec \phi_k = ( \phi_{0k}, \ldots, \phi_{N-1,k} ) \in \mathbb{ R }^N $ of $ P $ correspond to the values of an observable $ \phi_k : A \mapsto \mathbb{R } $ on the sampled states $ a_0, \ldots, a_{N-1} $; specifically, $ \phi_{nk} = \phi_k( a_n )$. In \cite{Giannakis16,DasGiannakis17} it was shown that as the number of delays $ q $ increases, $ \phi_k $ lies increasingly in a subspace of $ H $ corresponding to a single Koopman eigenspace. As a result,  the time series $ \vec \phi_k $ becomes increasingly monochromatic, with its frequency spectral power increasingly concentrated around a single frequency $ \omega_k $ in the Koopman spectrum. In other words, the kernel eigenfunctions obtained via this approach acquire a property that can be thought of as timescale separation; that is, they are able to decompose a broadband input signal ($x_n$) into a set of temporally coherent modes with frequencies intrinsic to the dynamical system. This kernel delay-embedding approach also improves robustness to observational noise \cite{Giannakis16}. As stated earlier, the key ingredient in the technique that ensures convergence to a single Koopman eigenspace is delay embeddings with many delays. A similar asymptotic behavior should hold for other classes of kernels besides the specific example in~\eqref{eqK}.     

While kernel eigenfunctions computed via this approach have been found useful for mode decomposition \cite{GiannakisMajda12a,BushukGiannakis15,SzekelyEtAl16,SzekelyEtAl16b,SlawinskaGiannakis2017} and prediction \cite{ZhaoGiannakis16,AlexanderEtAl17,ComeauEtAl17} in a variety of atmosphere ocean science applications, they can also be used in data-driven Galerkin schemes targeting Koopman eigenfunctions directly \cite{GiannakisEtAl15,Giannakis16,DasGiannakis17}. In effect, the kernel eigenfunctions obtained from delay-coordinate mapped data are already preconditioned to lie close to individual Koopman eigenspaces, and thus form a highly efficient basis for Koopman eigenfunction approximation.

For the purpose of computing Koopman eigenfunctions it is natural to consider the generator $ u $ of the Koopman group, defined as $ u( f ) = \frac{ d\ }{ dt } U_t f \rvert_{t=0} $. For measure-preserving systems, this operator is a skew-adjoint, unbounded operator defined on a dense domain $ D( u ) \subset H $ whose spectrum lies entirely on the imaginary line. Expressed in terms of $ u $, the Koopman eigenvalue problem reads $ u( z ) = i \omega z $; comparing this equation with~\eqref{eqUTEig} it is evident that the eigenvalues of $ U_t $ are exponentials of the eigenvalues of $ u $. 

As discussed in \cite{GiannakisEtAl15,Giannakis16,DasGiannakis17}, working with the bare generator $ u $ in data-driven schemes can be problematic for a number of reasons, including that it can have continuous spectrum and/or non-isolated eigenvalues. As a result, we solve instead the eigenvalue problem for the regularized generator,
\begin{displaymath}
  L z = \lambda z, \quad L = u - \eta \Delta,
\end{displaymath}      
where $ \eta $ is a positive number and $ \Delta $ a diffusion regularization operator constructed from the kernel operator $ P $; specifically, given $ f = \sum_k c_k \phi_k $, $ c_k \in \mathbb{ C }$, we set $ \Delta f = \sum_k \Lambda_k^{-1} c_k \phi_k $. Unlike $ u $, this operator always has discrete spectrum consisting of isolated eigenvalues \cite{FrankeEtAl10}. Galerkin methods can be used to numerically solve the eigenvalue problem for $ L $ in variatonal (weak) form, leading to a matrix generalized eigenvalue problem whose solutions give approximations to the  Koopman eigenfuncions in the form $ z = \sum_k c_k \phi_k $ \cite{Giannakis16,DasGiannakis17}. To order these eigenfunctions, we employ the Dirichlet energy functional $ E( z ) $ associated with $ \Delta $. In particular, we order the numerically computed Koopman eigenfunctions in order of increasing Dirichlet energy. This minimizes the risk of overfitting the training data since the eigenfunctions with the smallest Dirichlet energy are those exhibiting the least oscillatory behavior (with respect to the kernel $K $) on state space. 
     
\section{Selected applications to tropical climate science} 

In order to demonstrate the potential of our approach in the specific case of climate application we apply it to the high-dimensional datasets, with the particular emphasis on high-resolution Cloud Archive User Service (CLAUS) satellite infrared brightness temperature ($T_b$) data 
\cite{HodgesEtAl00} recorded over 23 years from July 1, 1983 to June 30, 2006. In the tropics, positive (negative) $ T_b $ anomalies are associated with reduced (increased) cloudiness, thus providing a surrogate for tropical convection. The data is sampled over the tropical belt from 15$^{\circ}$S to 15$^{\circ}$N with a resolution of 1$^{\circ}$ (in both longitude and latitude) generating 2D samples with $d_{\text{long}}=360$ longitude and $d_{\text{lat}}=31$ latitude gridpoints. We use the full 2D gridpoint $ T_b $ values arranged  prior to analysis into vectors of dimension $d = d_{\text{long}} \times d_{\text{lat}} =\text{11,160} $. Observations are collected at an interval of $\tau =3 $ h, producing a dataset with $s=\text{67,208}$ samples over the 23 years of the CLAUS record. The data contains the intensive observing period (IOP) of the Tropical Ocean Global Atmosphere Coupled Ocean Atmosphere Response Experiment (TOGA-COARE) which took place from November 1, 1992 to February 28, 1993.

\begin{figure}[ht]
\centering\includegraphics[width=1.05\linewidth]{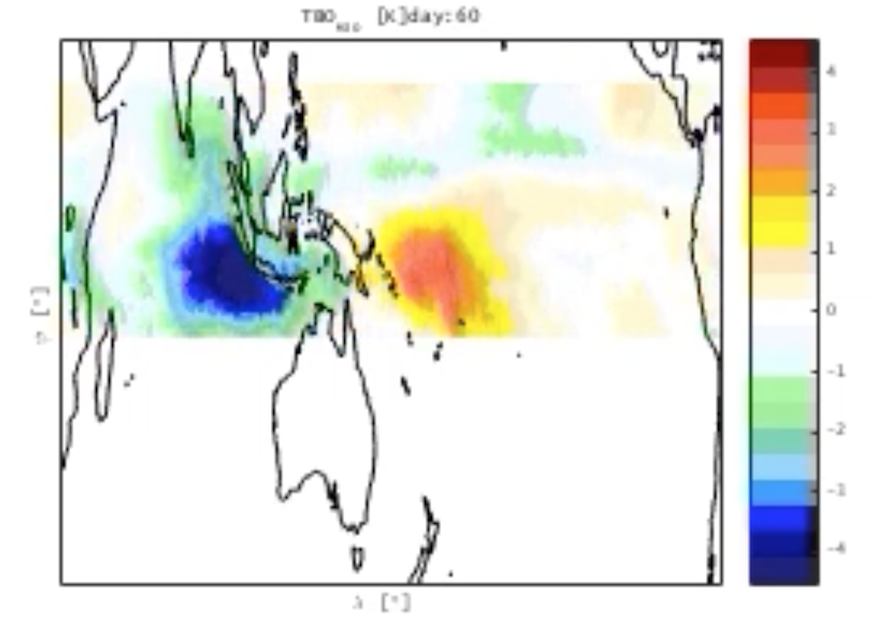}
\caption{Temporal snapshot of outgoing brightness temperature [K] projection onto spatio-temporal mode associated with intraseasonal oscillation (MJO) as obtained by our analysis of $\approx$ 25 years of outgoing brightness temperature sampled once per day.}
\end{figure}

The lag-embedded data $y_n$ are constructed using an intraseasonal embedding window of $\Delta t=64$ days. The number of lags corresponding to the $ \tau = 3 $ h sampling interval is $ q = \Delta t / \tau = 64 \times 8 = 512 $, meaning that the embedded data vectors populate a space of dimension $dq \approx 5.7 \times 10^6 $. In our analysis, we perform no preprocessing of the input data such as bandpass filtering or seasonal partitioning. 

Despite the fact that the ambient data space is high-dimensional, the system (i.e., tropical variability) at large scales is characterized by a relatively small number of patterns of interest. These patterns are well represented by eigenfunctions of the kernel operator $ P $ introduced earlier. In particular, we have extracted a hierarchy of modes of the tropical variability that span a wide range of timescales, such as the annual cycle and its harmonics, interannual (e.g., El-Ni\~no Southern oscillation, ENSO), intraseasonal, and diurnal modes \cite{SzekelyEtAl16, SzekelyEtAl16b}. One of our main results is the extraction of two distinct families of modes on the intraseasonal timescale -- the boreal winter Madden-Julian oscillation (MJO) and the boreal summer intraseasonal oscillation (BSISO). It is the first time that the two phenomena are identified as distinct modes by a data-driven method without preprocessing the input data by bandpass filtering and/or seasonal partitioning. These modes have different spatiotemporal patterns, i.e., an eastward-propagating pattern for MJO vs.\ a northeastward-propagating pattern for BSISO, and slightly different frequencies, i.e., $\sim 1/60$ days$^{-1}$ for MJO vs.\ $\sim 1/40$ days$^{-1}$ for BSISO. Our results are robust under changes of many input parameters, including temporal sparsity (e.g. sampling once per day, see Figure 1) or spatial dimension of the input signal.

\begin{figure}[ht]
\centering\includegraphics[width=1.05\linewidth]{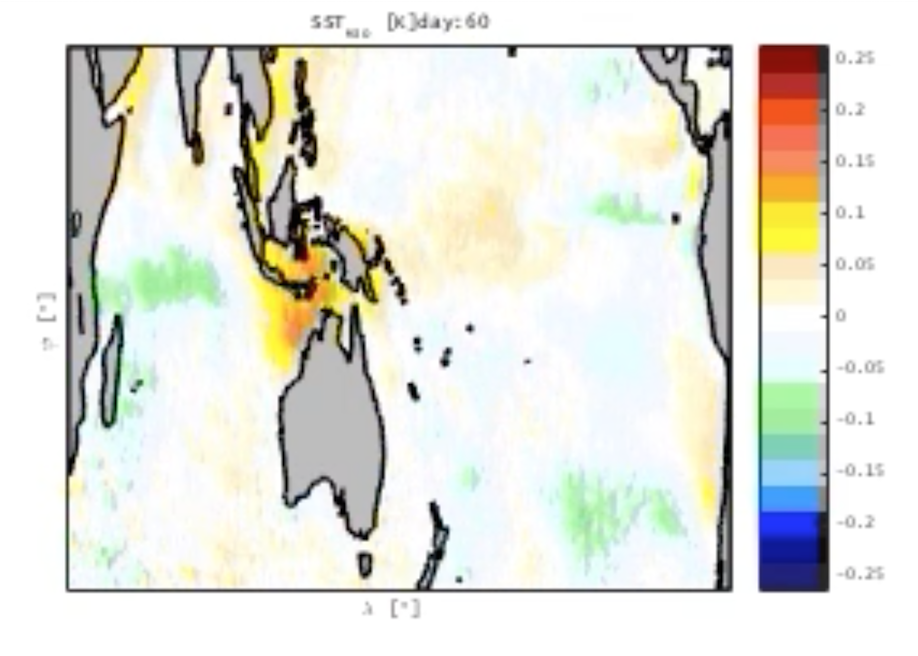}
\caption{Temporal snapshot of sea surface temperature [K]  projection onto spatio-temporal mode associated with MJO as extracted from $\approx$ 25 years of outgoing brightness temperature sampled once per day.}
\end{figure}

The ability to simultaneously detect different timescales of the system in a single process allows us to further study the interaction between them for additional analysis, e.g., predictability studies \cite{SzekelyEtAl16} and forecasting \cite{AlexanderEtAl17}. In \cite{SzekelyEtAl16} the leading kernel eigenfunctions were used as predictors to quantify the regime predictability of the MJO amplitude. Regimes are identified using a clustering method, and the information gain of each cluster is quantified using information-theoretic measures, such as relative entropy and mutual information. It is found that the most predictable MJO regimes occur in November and December, before the MJO active season which happens in February. A large part of the predictive information during the early season is explained by the ENSO interannual mode, and more precisely by El Ni\~no years with reemergence of predictability in January-February.  

Improving skillful extraction of MJO signals remains an important issue, as it projects strongly on climatic impacts across a range of spatiotemporal timescales. In particular, it is postulated that intraseasonal oscillations can dynamically lead to much slower oscillations on interannual timescales. In particular, ENSO phase changes are hypothesized to be caused by oceanic waves triggered by MJO. Here, we observe that our MJO eigenfunctions successfully recover the MJO-SST coupling (see Figure 1 and 2), with warm/cold SST anomaly preceding/following the MJO center of convection as expected \cite{DeMott14,DeMott15}. At the same time, other atmospheric-oceanic couplings possibly important for ENSO termination (e.g. enhanced westerlies in boreal spring over the western tropical Pacific Ocean  \cite{Stucker2013}) are well-recovered in datasets simulated by models that do not resolve many important features (e.g. MJO) observed in nature \cite{SlawinskaGiannakis2017,GiannakisSlawinska2017}.

\begin{figure}[ht]
\centering\includegraphics[width=1.05\linewidth]{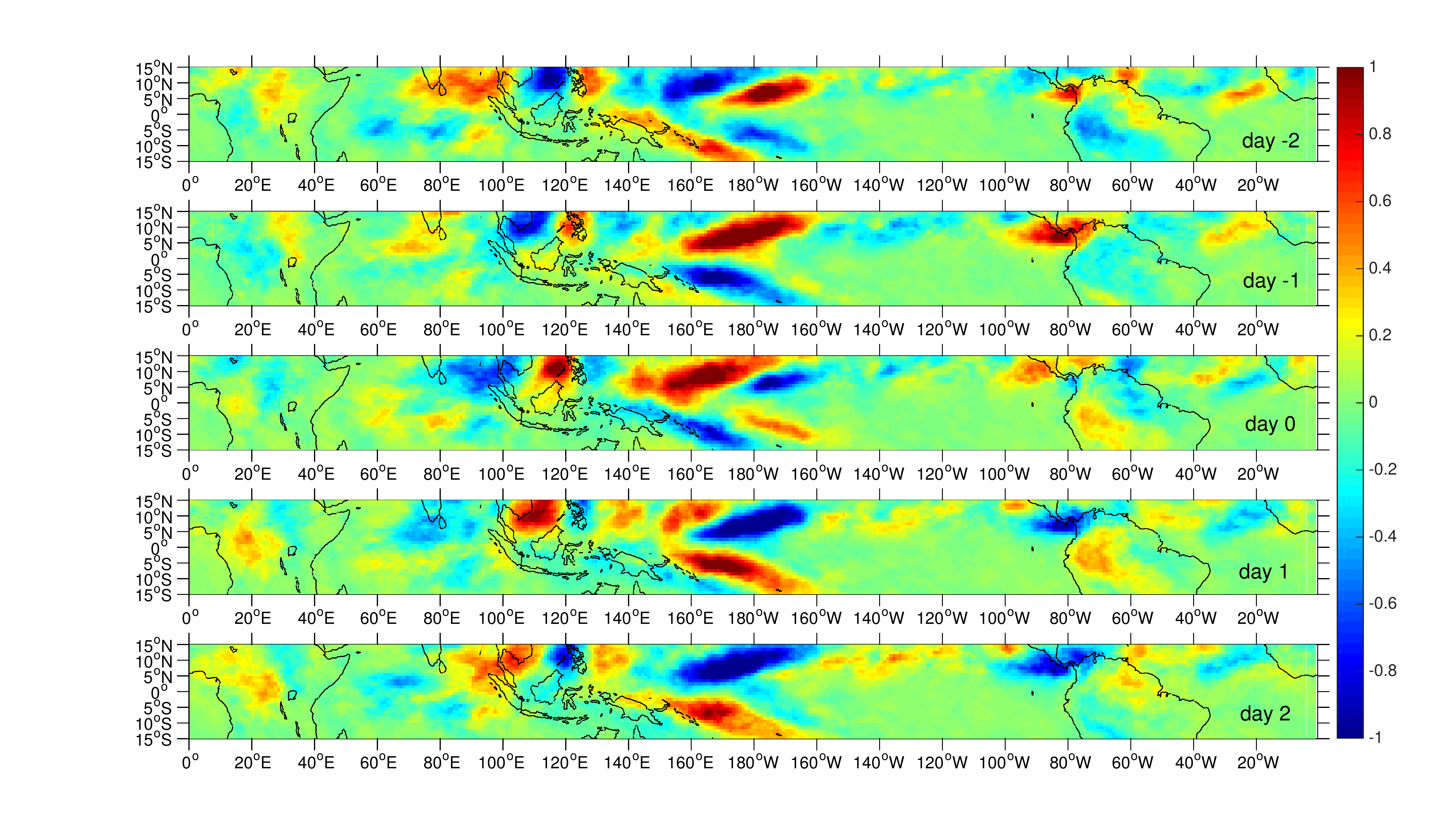}
\caption{5-day evolution of the cloud top temperature anomalies associated with the 5-day westward mixed Rossby-gravity wave, as discussed in \cite{SlawinskaGiannakis2016}.}
\end{figure}

Koopman eigenfunctions obtained via \cite{Giannakis16} approach are able to recover even cleaner spatiotemporal patterns of the above temporal scales, as well as propagating signals operating on faster (daily, see Figure 3)  timescales \cite{GiannakisEtAl15,SlawinskaGiannakis2016}.  These signals likely correspond to CCEWs \cite{KILADIS2009} predicted from idealized theories of atmospheric dynamics. Moreover, they are of smaller spatial scale than MJO that in turn does not have any theoretical correspondence and yet is hypothesized to be composed of CCEWs. As the statistical relation and physical mechanisms behind MJO interaction with larger and smaller spatiotemporal scales remain relatively poorly understood, our approach offers an attractive framework for further studies of the underlying nonlinear dynamics with the goal of addressing current problems in atmosphere ocean science.

\section{Summary}

This work demonstrates the potential of data-driven techniques in extracting dominant spatio-temporal modes from a high-dimensional datasets corresponding to sparse observations of complex dynamical systems. In particular, employing theoretical framework for dynamical systems and utilizing such data-driven metrics as kernels or Koopman operators \cite{Giannakis15, Giannakis16} allows one to identify several patterns that have a straightforward correspondence to variabilities that act over a range of timescales and are potentially associated with various  nonlinear interactions that impact dynamical evolution of the system. Moreover, novel nonparameteric extensions of these tools \cite{BerryEtAl15, Giannakis16, ZhaoGiannakis16} provide a range of data-driven approaches that allow to improve the performance of variety of predictive measures.

The Earth's tropical climate provide one example of dynamical system that can be studied with the above described techniques. They provide a new opportunity to study  evolution of tropical climate and in particular the role of nonlinear interactions across a range of spatio-temporal scales. Improved insight into physical nature of El Ni\~no/La Nin\~a (the most prominent variability of the Earth's climate that acts on interannual and planetary scales), and in particular seasonal locking of its emergence and termination, provides an example of observations that remain yet to be fully understood theoretically. Having here demonstrated skill in capturing many plausible components associated with this underlying phenomena, such as zonal propagation of atmospheric/oceanic waves triggered by MJO, or modulation of large-scale circulation scales by ENSO \cite{GiannakisEtAl15, GiannakisSlawinska2017, SlawinskaGiannakis2017}, allows one to examine their plausible/relative role in this regard. Moreover, having established refined approach in extracting tropical variabilities spanning a range of timescales, we hope to offer new schemes  \cite{SzekelyGiannakis2017} that potentially allow to improve climate prediction of probabilistic prediction of high-impact weather (e.g. extreme precipitation) \cite{Jones2004}.

\section{Acknowledgements}

This research was supported by DARPA grant HR0011-16-C-0116, NSF grant DMS-1521775, ONR grant N00014-
14-1-0150, ONR MURI grant 25-74200-F7112, and ONR YIP grant N00014-16-1-2649. The authors would like to thank Charlotte DeMott for useful discussion.

\end{document}